\documentclass[sloppy,oribibl,envcountsame,envcountsect,11pt]{article}

\usepackage[utf8]{inputenc} 
\usepackage{amsmath,amsthm,amssymb}
\usepackage{algorithm} 

\usepackage{newpxtext}
\usepackage[varg,bigdelims]{newpxmath}
\usepackage[margin=1in]{geometry}

\usepackage{comment}
\usepackage{float}

\usepackage{todonotes}

\title{Multivariate to Bivariate Reduction for Noncommutative
  Polynomial Factorization}

\author{V. Arvind\thanks{Institute of Mathematical Sciences, Chennai,
    India and Chennai Mathematical Institute, Siruseri, Kelambakkam,
    India, \texttt{email: arvind@imsc.res.in}} \and Pushkar S
  Joglekar\thanks{Vishwakarma Institute of Technology, Pune, India,
    \texttt{email: joglekar.pushkar@gmail.com}}}

\date{}

\newtheorem{theorem}{Theorem}[section]
\newtheorem{corollary}[theorem]{Corollary}
\newtheorem{definition}[theorem]{Definition}
\newtheorem{lemma}[theorem]{Lemma}
\newtheorem{fact}[theorem]{Fact}
\newtheorem*{claim}{Claim}
\newtheorem{remark}[theorem]{Remark}

\newtheorem{problem}[theorem]{Problem}
\newenvironment{proofof}[1]{\noindent{\it Proof of Claim. }}{{\bbox}}

  \def\bbox{\hspace*{\fill} $\blacksquare$\par\medskip}

  \DeclareMathOperator{\supp}{\text{supp}}

\usepackage{graphicx}

\newcommand{\op}[1]{\ensuremath{\operatorname{#1}}}

\newcommand{\M}{\mathcal{M}}

\newcommand{\F}{\mathbb{F}}
\renewcommand{\angle}[1]{\langle #1 \rangle}
\newcommand{\FX}{\F\angle{X}}

\newcommand{\Fxy}{\F\angle{x,y}}
\newcommand{\FXinfty}{\F\angle{X_{\infty}}}

\newcommand{\Q}{\mathbb{Q}}

\newcommand{\size}{\mathrm{size}}
\newcommand{\var}{\op{Var}}
\newcommand{\poly}{\mathrm{poly}}
\newcommand{\mon}{\op{Mon}}
\newcommand{\lm}{\op{lm}}

\usepackage{framed}

\begin{document}

\maketitle

\begin{abstract}
  Based on a theorem of Bergman \cite{Cohnrelations} we show that
  multivariate noncommutative polynomial factorization is
  deterministic polynomial-time reducible to the factorization of bivariate
  noncommutative polynomials. More precisely, we show the following:
  \begin{enumerate}
  \item In the white-box setting, given an $n$-variate noncommutative
    polynomial $f\in\FX$ over a field $\F$ (either a finite field or
    the rationals) as an arithmetic circuit (or algebraic branching
    program), computing a complete factorization of $f$ is
    deterministic polynomial-time reducible to white-box factorization of a
    noncommutative bivariate polynomial $g\in\F\angle{x,y}$; the
    reduction transforms $f$ into a circuit for $g$ (resp.\ ABP for
    $g$), and given a complete factorization of $g$ the reduction
    recovers a complete factorization of $f$ in polynomial time.

    We also obtain a similar deterministic polynomial-time reduction
    in the black-box setting.

  \item Additionally, we show over the field of rationals that
    bivariate linear matrix factorization of $4\times 4$ matrices is
    at least as hard as factoring square-free integers. This indicates
    that reducing noncommutative polynomial factorization to linear
    matrix factorization (as done in \cite{AJ22}) is unlikely to
    succeed over the field of rationals even in the bivariate case.
    In contrast, multivariate linear matrix factorization for $3\times
    3$ matrices over rationals is in polynomial time.
  \end{enumerate}
  
\end{abstract}

\section{Introduction}

The main aim of this paper is to show that multivariate polynomial
factorization in the free noncommutative ring
$\F\angle{x_1,x_2,\ldots,x_n}$ is polynomial-time reducible to
\emph{bivariate} noncommutative polynomial factorization in the
bivariate ring $\F\angle{x,y}$. Such a result for commutative
polynomial factorization is well-known due to Kaltofen's seminal work
on multivariate polynomial factorization in the commutative polynomial
ring $\F[y_1,y_2,\ldots,y_n]$ \cite{K89,KT90}. However, this problem
was open in the setting of noncommutative polynomials. Recently,
\cite{AJ22} a randomized polynomial-time algorithm was obtained for
the factorization of noncommutative polynomials over finite fields,
where the input polynomial is given by a noncommutative
formula.\footnote{Factorization of \emph{homogeneous} noncommutative
  polynomials is easier as it can be reduced to factorization of a
  special case of commutative polynomials. See \cite{ARJ15} for
  details.} Broadly speaking, the algorithm of \cite{AJ22} works via
Higman linearization (\cite{Hig} \cite{Cohnfir} \cite{GGOW20}) and
reduces the problem to linear matrix factorization which turns out to
have a randomized polynomial-time algorithm over finite fields.

\begin{problem}[Linear Matrix Factorization Problem]
The linear matrix factorization problem over a field $\F$ takes as
input a linear matrix: $L=A_0+\sum_{i=1}^n A_ix_i$, where the $A_i$
are $d\times d$ scalar matrices (over $\F$), the $x_i, 1\le i\le n$
are noncommuting variables, and $A_0$ is assumed invertible for
technical reasons. The problem is to compute a factorization of $L$ as
a product of irreducible linear matrices.
\end{problem}

The study of matrix factorization (linear matrix factorization, in
particular) is an important part of Cohn's factorization theory over
general free ideal rings. \cite{Cohnfir} \cite{Cohnrelations}.

Coming back to the polynomial factorization algorithm described in
\cite{AJ22}, the algorithm reduces polynomial factorization to linear
matrix factorization which is, in turn, reducible to the problem of
computing a common invariant subspace for a collection of $n$
matrices. The common invariant subspace problem can be efficiently
solved in the case of finite fields using Ronyai's algorithm
\cite{Ronyai} which is based on the Artin-Wedderburn theorem for
decomposition of algebras. This approach, however, runs into serious
difficulties when $\F$ is the field of rationals. The main difficulty
is that given a simple matrix algebra\footnote{i.e.\ the algebra has
  no nontrivial two-sided ideals.} $\mathcal{A}$ over rationals, we do
not know an efficient algorithm for finding out if $\mathcal{A}$ is a
division algebra or whether it has zero divisors. This is one of our
motivations for obtaining a reduction from multivariate polynomial
factorization to bivariate factorization. Because Higman Linearization
of a bivariate noncommutative polynomial given by a formula will yield
a bivariate linear matrix. One could hope that factorization of a
bivariate linear matrix is computationally easier than factorization
of an $n$-variate linear matrix. Unfortunately, this is not the
case. As we will see, even for $4$-dimensional bivariate linear
matrices the problem of factorization is at least as hard as factoring
square-free integers.

\paragraph{Multivariate to Bivariate}

We start with some formal preliminaries.  Let $\F$ be any field and
$X=\{x_1,x_2,\ldots,x_n\}$ be a set of $n$ free noncommuting
variables. Let $X^*$ denote the set of all free words (which are
monomials) over the alphabet $X$ with concatenation of words as the
monoid operation and the empty word $\epsilon$ as identity element.

The \emph{free noncommutative ring} $\FX$ consists of all finite
$\F$-linear combinations of monomials in $X^*$, where the ring
addition $+$ is coefficient-wise addition and the ring multiplication
$*$ is the usual convolution product. More precisely, let $f,g\in\FX$
and let $f(m)\in\F$ denote the coefficient of monomial $m$ in
polynomial $f$. Then we can write $f=\sum_m f(m) m$ and $g=\sum_m g(m)
m$, and in the product polynomial $fg$ for each monomial $m$ we have

\[
fg(m)=\sum_{m_1m_2=m} f(m_1)g(m_2).
\]

The \emph{degree} of a monomial $m\in X^*$ is the length of the
monomial $m$, and the degree $\deg f$ of a polynomial $f\in\FX$ is the
degree of a largest degree monomial in $f$ with nonzero coefficient.
For polynomials $f,g\in\FX$ we clearly have $\deg (fg) = \deg f + \deg
g$.

A \emph{nontrivial factorization} of a polynomial $f\in\FX$ is an
expression of $f$ as a product $f=gh$ of polynomials $g,h\in\FX$ such
that $\deg g > 0$ and $\deg h > 0$. A polynomial $f\in\FX$ is
\emph{irreducible} if it has no nontrivial factorization and is
\emph{reducible} otherwise.  For instance, all degree $1$ polynomials
in $\FX$ are irreducible. Clearly, by repeated factorization every
polynomial in $\FX$ can be expressed as a product of
irreducibles.

The problem of noncommutative polynomial identity testing (PIT) for
multivariate polynomials is known to easily reduce to noncommutative
PIT for bivariate polynomials: the reduction is given by the
substitution
\[
x_i\to xy^i, 1\le i\le n,
\]
which transforms a given arithmetic circuit (or formula or algebraic
branching program) computing a polynomial $f(x_1,x_2,\ldots,x_n)$ to
the bivariate polynomial $g(x,y)=f(xy,xy^2,\ldots,xy^n)$. As this
substitution map ensures that every monomial of $f$ is mapped to a
distinct monomial of $g(x,y)$, it easily follows that $f$ is the zero
polynomial if and only if $g(x,y)$ is the zero polynomial.

Indeed, it can be shown \cite[Excercise 2.5, Problem 14]{Cohnfir} that
this substitution map give an injective homomorphism from the ring
$\F\angle{x_1,x_2,\ldots,x_n}$ to $\F\angle{x,y}$. Unfortunately, this
map does not preserve factorizations. For example, the polynomial $f =x_3x_1+x_4x_2+x_4x_1+x_5x_2 \in \FX$ is clearly irreducible but image of $f$ under the above map non trivially factorizes as $(xy^2 + xy^3)(yxy + y^2xy^2)$. Thus, we cannot use this substitution map to
obtain a reduction from noncommutative multivariate polynomial
factorization to bivariate polynomial factorization.


\paragraph{Bergman's 1-inert embedding}

However, based on a theorem of Bergman \cite[Chapter
  4]{Cohnrelations}, we can obtain a polynomial-time reduction from
factorization of multivariate noncommutative polynomials in $
\F\angle{x_1,x_2,\ldots,x_n}$ given by arithmetic circuits (respect. noncommutative algebraic
branching programs(ABP)) to factorization of bivariate noncommutative
polynomials in $\F \langle x, y \rangle$, again given by arithmetic circuit (respect. an ABP). This
reduction is polynomial-time bounded for both finite fields and
rationals. In the case of rationals we need to ensure that the bit
complexities of all numbers involved are polynomially
bounded. Furthermore, we show that essentially the same reduction
works in the black-box setting as well.

More precisely, Bergman's theorem \cite[Chapter 4, Theorem
  5.2]{Cohnrelations} shows a 1-inert embedding of free algebras of
countable rank into free algebras of rank 2. The property of
1-inertness of the embedding map is defined below. We restrict the
definition to free noncommutative polynomial rings.

\begin{definition}[1-inert embedding]{\rm\cite{Cohnrelations}}\label{inert-def}
  Let $X=\{x_1,x_2,\ldots\}$ be a countably infinite set of free
  noncommuting variables and $Y=\{x,y\}$ be two free noncommuting
  variables.  A \emph{1-inert embedding} of $\F\angle{X}$ into
  $\F\angle{Y}$ is an injective homomorphism $\varphi:\F\angle{X}\to
  \F\angle{Y}$ such that for each polynomial $f\in\F\angle{X}$, if its
  image $\varphi(f)$ factorizes nontrivially in $\F\angle{Y}$ as
    \[
    \varphi(f) = g_1\cdot g_2,
    \]
    then their preimages $\varphi^{-1}(g_1)$ and $\varphi^{-1}(g_2)$
    exist and, since $\varphi$ is a homomorphism, it gives a
    nontrivial factorization $f=\varphi^{-1}(g_1)\varphi^{-1}(g_2)$ of
    $f$ in the ring $\F\angle{X}$.
\end{definition}    

\begin{remark}
  The above definition implies that for all factorizations
  $\varphi(f)=g_1 g_2$, the polynomials $g_1$ and $g_2$ are in the
  range of $\varphi$. We note that Cohn's work
  \cite{Cohnfir,Cohnrelations} treats 1-inert embeddings $\varphi: R_1\to
  R_2$ for general noncommutative integral domains $R_1$ and $R_2$. In
  the general case, the definition only requires that there is some
  \emph{unit} $u\in R_2$ such that $g_1u$ and $u^{-1}g_2$ are in the
  range of the map $\varphi$ which means that the factorization
  $\varphi(f)=(g_1u)(u^{-1}g_2)$ holds in the range of $\varphi$.  As
  the only units in $\FX$ are elements of $\F$, for a polynomial $f
  \in \F\angle{X}$, the factors of its image $\varphi(f)$ are all in
  the range $\varphi(\F\angle{X})$ of the 1-inert embedding $\varphi$.
\end{remark}

\begin{definition}
  A \emph{complete factorization} of noncommutative polynomial
  $f\in\FX$ is a factorization $f=f_1\cdot f_2\cdots f_r$ into a
  product of irreducible polynomials $f_i\in\FX$.
\end{definition}

Given an algebraic branching program (respec. Arithmetic Circuit) for
$f$, we can efficiently obtain an algebraic branching program
(respec. Arithmetic Circuit) for $\varphi(f)$ and then we use idea of
running a substitution automata on ABPs or circuits (see
e.g. \cite{ARJ15}, \cite{AJMR17}, \cite{AMS10}) to construct a
complete factorization of $f$ given a complete factorization of
$\varphi(f)$. In the next section we will elaborate and expand upon
Bergman's embedding theorem \cite{Cohnrelations} and show how to get
its effective algorithmic version which is useful for our purpose of
reconstruction of factors of $f$ from factors of $\varphi(f)$.

\section{Bergman's embedding}

We define a total ordering $\prec$ on monomials in $\{x,y\}^*$ as follows. 
\begin{definition}
For $m_1,m_2 \in \{x,y\}^*, m_1 \neq m_2$, we say \emph{$m_1 \prec
  m_2$} if one of the following holds:
\begin{itemize}
\item $m_1=m_2$.  
\item $\deg(m_1)< \deg(m_2)$.
\item $\deg(m_1)=\deg(m_2)$ and if $m_1\ne m_2$ then the leftmost position
$i$ where they differ we have $m_1[i]=y$ and $m_2[i]=x$.
\end{itemize}  
\end{definition}

The above ordering is just the usual lexicographic ordering on binary
strings treating $y$ as $0$ and $x$ as $1$. For any polynomial $g$,
let $\mon(g)$ denotes set of all monomials of $g$ with non-zero
coefficient.

When $m_1 \prec m_2$ we say that monomial $m_1$ is \emph{smaller} than
monomial $m_2$. Equivalently, $m_2$ is \emph{larger} than $m_1$.  The
\emph{leading monomial} of a polynomial $g \in \F\langle x, y\rangle$
is the monomial $m \in \mon(g)$ (denoted by $\lm(g)$) such that $w
\prec m$ for all $w\in\mon(g)$. That is, the leading monomial of $g$
is the largest monomial in $\mon(g)$.
 

\begin{definition}
For a monomial $m \in \{x, y\}^*$ let $d_x(m)$ (respectively,
$d_y(m)$) denote the number of occurrences of $x$ (respectively, $y$)
in $m$. The \emph{imbalance} $i(m)$ of the monomial $m$ is defined
as
\[
i(m)=d_x(m)-d_y(m).
\]
\end{definition}

Let $B$ be the algebra of all polynomials such that every monomial of
the polynomials has imbalance $0$.

Let $T$ be set of all monomials $m$ such that either $m = \epsilon$ or
$i(m)=0$ and for any prefix $m'$ of $m$ with $m' \neq \epsilon$, $m'
\neq m$, $i(m')>0$. We call these monomials as \emph{minimally
  balanced monomials}. Clearly in all the non-empty monomials in $T$
the leftmost symbol is $x$. We arrange the non-empty monomials in $T$
in lexicographic ordering $\prec$ and the $i^{th}$ monomial in the
sequence is denoted by $u_i$. Let $\overline{u_i}$ is a monomial
obtained by replacing every occurrence of $x$ by $y$ and $y$ by $x$ in
$u_i$. Let $\overline{T} = \{\overline{u_i} ~|~ i\geq 1 \}$. It is
clear that the monomials in $T$ and $\overline{T}$ together generate
the algebra $B$.


Let $C$ be an algebra generated by $\{u_i + \overline{u_i} ~|~ i \geq
1\}$. Clearly $C$ is a subalgebra of $B$.


As each $u_i$ and $\overline{u_i}$ is balanced, clearly the imbalance
of any monomial of a polynomial in $C$ is $0$. Let $X= \{x_1, x_2,
\ldots, x_n\}$ and $X_{\infty} =\{x_1, x_2, \ldots \}$ be a countably
infinite set of indeterminates.

We observe a crucial property of polynomials in $B \setminus C$.

\begin{lemma}\hfill{~} \label{lemma-bc}
\begin{itemize}
\item The leading monomial $m$ of any polynomial in $C$ has the form
  $m = u_{i_1}u_{i_2}\cdots u_{i_\ell}$, where each $u_{i_j}\in
  T$. That is, $m$ does not have as subword any
  $\overline{u}\in\overline{T}$.
\item Every polynomial $f \in B \setminus C$ can be expressed as
  $f=g+h$ for $g\in C$ and $h\in B$. Moreover, if $h\ne 0$ then the
  leading monomial of $h$ has some $\overline{u}\in\overline{T}$ as
  subword.
\end{itemize}
\end{lemma}

\begin{proof}
By definition of algebra $B$ (and hence $C$), for $g\in B$ and any
monomial $m\in\supp(g)$ we have $m\in (T \cup
\overline{T} )^*$.  Moreover, each $g\in C$ is an linear combination
of products of the form $\prod_{k=1}^\ell(u_{i_k} +
\overline{u_{i_k}})$. Hence, if $\deg(g)=d$ and
$\supp(g)$ contains a degree-$d$ monomial $g_{j_1} g_{j_2} \ldots
g_{j_\ell}$, where $g_{j_k} \in \{ u_{j_k}, \overline{u_{j_k}} \}$ for
$k \in[\ell]$, then $\supp(g)$ also contains the degree-$d$ monomial
$u_{j_1} u_{j_2} \ldots u_{j_\ell}$. By the definition $\prec$, the
monomial $u_{j_1} u_{j_2} \ldots u_{j_\ell}$ is larger than (with respect to ordering $\prec$) all the monomials with some
$\overline{u}\in\overline{T}$ as a subword. Therefore, the leading
monomial of any polynomial $g\in C$ has the form claimed.

Next, let $f \in B \setminus C$. If the leading monomial of $f$ has a
subword $\overline{u}\in\overline{T}$ then the claim follows as $f =
0+f$ and $0 \in C$. Suppose the leading monomial of $f$ is $m=u_{j_1}
u_{j_2} \cdots u_{j_\ell}$, $u_{j_k} \in T$ for all $k$.  If
coefficient of $m$ in $f$ is $\alpha$, Let 
\begin{equation}\label{repeat-subtract}
f_1 = f- \alpha (u_{j_1}+ \overline{u_{j_1}})(u_{j_2}+
\overline{u_{j_2}})\ldots (u_{j_\ell}+ \overline{u_{j_\ell}}).
\end{equation}
If $m_1$ is the leading monomial of $f_1$ then clearly $m_1 \prec m$.
Furthermore, $f_1\in B\setminus C$ as $f-f_1\in C$. Hence, it suffices
to show $f_1=g_1+h_1$ for some $g_1\in C$ and $h_1\in B$ with the
claimed property. We can apply the subtraction step of
Equation~\ref{repeat-subtract} to $f_1$ to obtain $f_2$ and so on,
where in the $i^{th}$ step we obtain a new polynomial $f_i$ whose
leading monomial is smaller than the leading monomials of $f$ and
each $f_j, j<i$. Since the $\prec$-ordering is a well-ordering on
monomials, this process will terminate giving us the desired
expression of $f$ as $f=g+h$ where $g \in C$ and the leading term of
$h$ has a subword $\overline{u}$ for some $u\in T$. This proves the
second part.
\end{proof}

\begin{lemma}
There is an injective homomorphism (i.e.\ a homomorphic embedding)
from the ring $\FXinfty$ to $\Fxy$.
\end{lemma}

\begin{proof}
Consider the function $\varphi:\FXinfty \mapsto \Fxy$ defined as
follows:

\begin{itemize}
\item Let $\varphi(x_i) = u_i + \overline{u_i}$ for all $x_i\in X_\infty$.
\item Extend $\varphi$ to all monomials by multiplication. That is,
  $\varphi (x_{i_1}x_{i_2}\ldots x_{i_k}) = \prod_{j =1}^k \varphi(x_{i_j})$.
\item Further, extend $\varphi$ to the ring $\F \langle X_{\infty}
  \rangle$ by linearity: $\varphi( \sum_{i=1}^t \alpha_{i} m_i ) =
  \sum_{i=1}^t \alpha_i \varphi(m_i)$, for monomials $m_i \in
  X_{\infty}^*$ and scalars $\alpha_i \in \F$ for $i=1$ to $t$.
\end{itemize}

To see that $\varphi$ is a homomorphism, we first note that, by
linearity, we have $\varphi(f+g)= \varphi(f)+\varphi(g)$ for $f,g\in
\FXinfty$.  To verify that $\varphi(fg)=\varphi(f)\varphi(g)$, let $f
= \sum_m f_m m$ and $g=\sum_m g_m m$ where $f_m, g_m\in \F$ are the
coefficients of monomial $m$ in $f$ and $g$, respectively. Then
\begin{eqnarray*}
\varphi(fg) &=& \varphi\left( (\sum_m f_m m)(\sum_w g_w w)\right)\\
&=& \varphi\left(\sum_{m,w} f_m g_w mw\right)\\
&=& \sum_{m,w}f_m g_w \varphi(mw) ~~~~(\text{by linearity of }\varphi)\\
&=& \left(\sum_m f_m \varphi(m)\right)\left(\sum_w f_w \varphi(w)\right)\\
&=& \varphi(f) \varphi(g).
\end{eqnarray*}

In order to show $\varphi$ is injective, it suffices to show
$\varphi(f)\ne 0$ for $f\ne 0$. Suppose $m\in \supp(f)$. Then we note
that $\varphi(m)\ne 0$ by the definition of $\varphi$. Hence, if $m$
is the only monomial in $\supp(f)$ it follows that $\varphi(f)\ne 0$.

Otherwise, let $m'\in\supp(f)$ and $m'\ne m$. Let $u$ be largest
common prefix of $m$ and $m'$. Then
\[
m = ux_iv \text{ and } m'=ux_jw,
\]
for monomials $u,v,w \in X_\infty^*$ and $x_i \neq x_j$. Noting that
$\varphi(x_i) =u_i+\overline{u}_i$ and $\varphi(x_j)=u_j+\overline{u}_j$
we have
\[
\varphi(m)=\varphi(u)(u_i+\overline{u}_i)\varphi(v) \text{ and }
\varphi(m')=\varphi(u)(u_j+\overline{u}_j)\varphi(w).
\]
From the definition of $\varphi$, clearly $\varphi(u)$ is a
homogeneous polynomial in $\Fxy$. Let $\deg(\varphi(u))=D$.  Suppose
$\ell = |u_i|=|\overline{u}_i|$ and $\ell' =
|u_j|=|\overline{u}_j|$. We can assume without loss of generality that
$u_i\prec u_j$. Hence $\ell\le \ell'$. As $u_i$ and $u_j$ are
minimally balanced, $u_i$ cannot be a prefix of $u_j$. Therefore, for
any monomials $w_1, w_2$ in $\supp(\varphi(m_1))$ and
$\supp(\varphi(m_2))$, respectively, $w_1$ and $w_2$ will differ in
the length $\ell$ subword starting at location $D+1$. It follows that
$\supp(\varphi(m))\cap(\varphi(m')) =\emptyset$. Hence, $\varphi(f)\ne
0$ implying that $\varphi$ is injective.
\end{proof}


We next have an important property about factorization of polynomials
in the algebra $C$. In order to keep our presentation self-contained
we give a complete proof with more details than in Cohn's book
\cite{Cohnrelations}.


\begin{theorem}[Bergman] \cite[Chapter 4, Theorem 5.2]{Cohnrelations}\label{thm-bergman}
Let $f \in C$. For any factorization $f= g \cdot h$ the polynomials
$g$ and $h$ are in $C$.
\end{theorem}

\begin{proof}
First we show that all monomials of $g$ have the same
imbalance. Likewise, all monomials of $h$ have the same
imbalance. Suppose $a_{min}$ and $a_{max}$ are the minimum and the
maximum imbalances of monomials of $g$. Let $b_{min}$ and $b_{max}$ be
the minimum and the maximum imbalance of monomials of $h$. Let
$m_{min}$ be a smallest monomial (with respect to $\prec$) among all monomials of $g$ with
imbalance $a_{min}$, and $m_{max}$ be a largest monomial (with respect to $\prec$) among all the monomials of $g$ with imbalance $a_{max}$. Let $w_{min},w_{max}$ be monomials similarly
defined for polynomial $h$ corresponding to $b_{min}$ and
$b_{max}$. Now consider the monomial $u= m_{max} w_{max}$. It is
non-zero in $f=g. h$ and has imbalance $a_{max}+b_{max}$. Similarly,
monomial $v=m_{min} w_{min}$ is non-zero in $f$ and has imbalance
$a_{min}+b_{min}$. As $f\in C \subset B$, each monomial of $f$ has
imbalance $0$. Hence, $a_{max}+b_{max}=0$ and $a_{min}+b_{min}=0$. So
$a_{max}= - b_{max} \leq -b_{min} = a_{min}$, implying
$a_{min}=a_{max}=a$ and $b_{min}=b_{max}=-a$. Thus, all monomials of
$g$ have imbalance $a$ and all monomials of $h$ have imbalance $-a$.

Let $m$ be the leading monomial of $f$. Clearly, $m$ is a maximum degree
monomial of $f$. Moreover, $m$ is largest among the
max-degree monomials of $f$. Let $m =m_1 m_2$ with $m_1 \in \supp(g)$
and $m_2 \in \supp(h)$. We have $i(m_1)=a$, $i(m_2)= -a$. As $f\in C$,
the monomial $\bar{m}$ obtained by replacing every occurrence of $x$
by $y$, and $y$ by $x$ in $m$ is also in $\supp(f)$.  Moreover,
$\bar{m}$ is the smallest monomial among the
max-degree monomials of $f$. This forces that the monomial $\bar{m}_1$
(obtained by interchanging $x,y$ in $m_1$) is in
$\supp(g)$. Similarly, monomial $\bar{m}_2$ (obtained by swapping
$x,y$ in $m_2$) is in $\supp(h)$. We have $i(\bar{m_1})= -a$ and
$i(\bar{m_2})=a$. Now, all the monomials of $g$ have the same
imbalance, and $m_1, \bar{m}_1\in\supp(g)$. This forces
$a=-a=0$. Consequently, all monomials in $\supp(g)\cup \supp(h)$ have
imbalance zero which implies $g, h\in B$.

By Lemma \ref{lemma-bc} applied to $g$ and $h$ we have
\begin{enumerate}
\item $g=g_1+g_2$, $h=h_1+h_2$, $g_1,h_1\in C$, $\lm(g_2)$ contains
  $\bar{u}\in\bar{T}$, and $\lm(h_2)$ contains $\bar{v}\in\bar{T}$,
\item Consequently, the $\deg(g_2)$ prefix of $\lm(g_2h_1)$ contains
  the subword $\bar{u}$ and the $\deg(h_2)$ suffix of $\lm(g_1h_2)$
  contains the subword $\bar{v}$.
\item Finally, the $\deg(g_2)$ prefix and the $\deg(h_2)$ suffix of
  $\lm(g_2\cdot h_2)$ contains both subwords $\bar{u}$ and $\bar{v}$.
\end{enumerate}
 
Hence the leading monomials $\lm(g_2\cdot h_1), \lm(g_1 \cdot h_2)$, and
$\lm(g_2 \cdot h_2)$ cannot cancel with each other. As a consequence,
the leading monomial of $g_2\cdot h_1 + g_1 \cdot h_2 + g_2 \cdot h_2$
contains a sub-word from $\bar{T}$ unless both $g_2=0$ and $h_2
=0$. Hence,
\begin{eqnarray*}
  g_2\cdot h_1 + g_1 \cdot h_2 + g_2 \cdot h_2 & =& gh - g_1 \cdot
  h_1 \in C \text{ and}\\
  f & = & g\cdot h , g_1, h_1 \in C.
\end{eqnarray*}

By Lemma \ref{lemma-bc}, for any polynomial $\frak{f}\in C$ its leading
monomial $\lm(\frak{f})$ cannot have a subword from $\bar{T}$. It
forces $g_2=0$ and $h_2=0$ which implies $g,h\in C$.
\end{proof}

The following theorem, which is a consequence of
Theorem~\ref{thm-bergman} shows that the embedding $\varphi$ is a
1-inert embedding (see Definition~\ref{inert-def}). That is, it
preserves factorizations.

\begin{theorem}\label{thm-inert}
Let $f \in \FX$, where $X=\{x_1, \ldots, x_n\}$. Suppose $f'=
\varphi(f)= g' \cdot h'$ is a non-trivial factorization of
$\varphi(f)$ in the ring $\F\angle{x,y}$. Then there exist polynomials
$g, h \in \FX$, $g,h \not\in \F$ such that $g' = \varphi(g)$, $h'=
\varphi(h)$ and $f = g \cdot h$.
\end{theorem}

\begin{proof}
  By construction, the homomorphism $\varphi$ injectively maps
  $\FXinfty$ into $\F\angle{x,y}$. As $\FX\subset \FXinfty$, $\varphi$
  maps $f \in \FX$ to some $f'=\varphi(f) \in C$. Suppose $f'=g'\cdot
  h'$ is a nontrivial factorization of $f'$ in $\F\angle{x,y}$. By
  Theorem \ref{thm-bergman}, as $f'\in C$ both the factors $g',h'\in
  C$.  Since $g'\in C$, it is an $\F$-linear combination of products
  of the form $(u_{t_1}+ \overline{u_{t_1}}) (u_{t_2}+
  \overline{u_{t_2}}) \ldots (u_{t_{\ell}}+
  \overline{u_{t_\ell}})$. By definition of $\varphi$,
 \[
  (u_{t_1}+ \overline{u_{t_1}})(u_{t_2}+ \overline{u_{t_2}}) \ldots
  (u_{t_{\ell}}+ \overline{u_{t_\ell}}) =
  \varphi(x_{t_1}x_{t_2} \ldots x_{t_\ell}).
\]
Hence, by linearity, it follows that $g' = \varphi(g)$ for some
nontrivial polynomial $g \in \FXinfty$, similarly there is a
nontrivial polynomial $h \in \FXinfty$ such that $h' =
\varphi(h)$. Since $\varphi$ is a homomorphism, we have
\[
\varphi(f) =f'=g'\cdot h'= \varphi(g) \cdot\varphi(h)=\varphi(g\cdot h).
\]
As $\varphi$ is injective, we have $f=g\cdot h$.  To complete the
proof we need to argue that $g, h \in \FX$. Let $\var(g)$ denotes set
of variables $x_i$ which appears in some non-zero monomial of $g$. We
want to show that $\var(g) \subseteq X$. Suppose $\var(g)$ contains
some $x_i \not \in X$.  Among all monomials of $g$ containing $x_i$,
let $m$ be the largest monomial (under $\prec$-ordering). Then the
monomial $m \cdot \lm(h)$ contains the variable $x_i$ and has a
non-zero coefficient in $f=gh$. This is a contradiction as $f \in \FX$
and $X$ does not contain $x_i$. Hence $\var(g)\subseteq X$. Similarly,
$\var(h)\subseteq X$.
\end{proof}


\section{Multivariate to Bivariate reduction} \label{sec-multi-bi}

In this section we will apply Bergman's theorem to show that
multivariate noncommutative polynomial factorization is reducible to
bivariate noncommutative polynomial factorization in both white-box
and black-box.

We first describe some simple tools using which we can obtain an
efficient reduction from Bergman's theorem (Theorem~\ref{thm-bergman}).

Let $X=\{x_1, x_2, \ldots, x_n\}$, and $v_1, v_2, \ldots, v_n$ be any
$n$ distinct and minimally balanced monomials in $\{x,y\}^*$. We
define function $\varphi:\FX \to \Fxy$:
\begin{itemize}
\item $\varphi(x_i) = v_i + \overline{v_i}$ for all $i$.
\item $\varphi$ is extended to monomials by multiplication,
  i.e. $\varphi (x_{i_1}x_{i_2}\ldots x_{i_k}) = \prod_{j =1}^k
  \varphi(x_{i_j})$. 
\item $\varphi$ is extended to $\F \langle X \rangle$ by linearity.
\end{itemize}

\begin{remark}
  The above definition is essentially like in the proof of Bergman's
  theorem, except that here $X$ is a finite set of variables and the
  $v_i, 1\le i\le n$ are any $n$ distinct minimally balanced
  monomials.
\end{remark}

We can show the following along the same lines as
Theorem~\ref{thm-bergman} and Theorem~\ref{thm-inert}.  The
straightforward proof is by a suitable renaming of the variables $x_1,
\ldots, x_n$ before and after application of Theorem \ref{thm-bergman}
in the proof of the Theorem \ref{thm-inert}.

\begin{lemma}\label{lemma-inert}
Let $X=\{x_1, \ldots, x_n\}$, $f \in \FX$. Suppose $v_1, v_2, \ldots,
v_n$ are any distinct minimally balanced monomials in $\{x,y\}^*$. If
$f'=\varphi(f)= g' \cdot h'$ is a non-trivial factorization of $f'$ in
$\F\angle{x,y}$ then there are polynomials $g, h \in \FX$ such that
$g' = \varphi(g)$, $h'= \varphi(h)$ and $f = g \cdot h$.
\end{lemma}



In order to obtain polynomial-time computable reduction it is
convenient to choose $v_1, v_2, \ldots, v_n$ such that each $v_i$ has
the same length. The next lemma ensures that $\ell=O(\log n)$
suffices.

\begin{lemma}\label{lemma-catlan}
  There are at least $n$ minimally balanced monomials of length
  $2\ell$ in $\{x,y\}^*$ for $\ell \geq \text{max} (\lceil \log 4n
  \rceil, 7)$. Furthermore, the lexicographically first $n$ minimally
  balanced monomials of length $2\ell$ can be computed in time
  polynomial in $n$.
\end{lemma}

\begin{proof}
First we consider the number of minimally balanced monomials of length
$2\ell$ for $\ell\geq 2$. The first symbol of any minimally balanced
monomial is $x$. If it is more than $2$, the second symbol is also
$x$ (if it was $y$, then the balanced monomial $xy$ would be a strict
prefix of the minimally balanced monomial, which is a contradiction.)
We consider monomials of the form
\[
v = xx\cdot w \cdot yy,
\]
where $w$ is a Dyck monomial\footnote{Essentially a balanced
  parenthesis string with $x$ as left and $y$ as right parenthesis,
  respectively}. That is, $w$ is a balanced monomial such that every
prefix of $w$ has at most as many $y$'s as $x$'s. Notice that $w\in
\{x, y\}^{2\ell-4}$. It follows that any nontrivial prefix of $v$ has
strictly more $x$ than $y$. So any such monomial is minimally balanced
of length $2\ell$. The number of Dyck monomials of length $2\ell-4$ is
$C_{\ell-2}$ (the $(\ell-2)^{th}$ Catalan number). A standard
estimate yields
\[
C_k \sim \frac{4^k}{k^{3/2} \sqrt{\pi}},
\]
which implies that $C_k$ is $2^{\Omega(k)}$. Specifically, $C_k > 2^k$
for $k \geq 5$. If $n < 2^{\ell-2}$ and $\ell\geq 7$ then there are at
least $n$ minimally balanced monomials of length $2\ell$, for $\ell =
\text{max} (\lceil \log 4n \rceil, 7)$.

Clearly, we can compute the $v_i, 1\le i\le n$ by enumeration in
$\poly(n)$ time.
\end{proof}

\subsection{White-box reduction}

We first describe the reduction in the white-box case for input
polynomial $f\in\FX$ given by a noncommutative arithmetic circuit.

\begin{lemma}
Let $X=\{x_1, \ldots, x_n\}$ and $f\in \FX$ be a noncommutative
polynomial given by arithmetic circuit $C$ of size $s$. Then there is
a deterministic polynomial time algorithm that outputs an arithmetic
circuit computing the polynomial $\varphi(f)\in\F\angle{x,y}$, where
the minimally balanced monomials $v_i, 1\le i\le n$ defining the map
$\varphi$ are as described by Lemma~\ref{lemma-catlan}.
\end{lemma}


\begin{proof}

For $1\le i\le n$, we note that the sum of two monomials
$v_i+\overline{v_i}$ can be computed by a noncommutative arithmetic
formula $F_i$ of size $O(\log n)$. Let $C'$ be the arithmetic circuit
obtained from circuit $C$ by replacing input variable $x_i$ with the
formula $F_i$. Clearly, $C'$ computes $\varphi(f)$ and its size
is polynomially bounded.
\end{proof}

\begin{lemma}
  For $f\in\FX$ suppose $\varphi(f)=f_1'\cdot f_2'\cdots f_r'$ is a
  complete factorization of $\varphi(f)$ in $\F\angle{x,y}$ into
  irreducible factors $f_i'\in\F\angle{x,y}$. Then there are
  irreducible polynomials $f_1, f_2, \ldots, f_r\in\FX$ such that
  $f=f_1f_2\ldots f_r$ and $\varphi(f_i)=f_i'$ for each $i$.
\end{lemma}

\begin{proof}
It follows by repeated application of Lemma \ref{lemma-inert} that if
\[
\varphi(f)= f_1'\cdot f_2'\cdots f_r',
\]
is a factorization into irreducible factors $f_i'\in\F\angle{x,y}$,
then there are polynomials $f_1, f_2, \ldots, f_r\in\FX$ such that
$f=f_1f_2\ldots f_r$ and $\varphi(f_i)=f_i'$ for each $i$. We claim
each $f_i$ is irreducible. For, if $f_i = g \cdot h$ is a nontrivial
factorization of $f_i$ in $\FX$ then clearly $f_i' =
\varphi(f_i)=\varphi(g)\varphi(h)$ is a nontrivial factorization of
$f_i'$, which contradicts its irreducibility.
\end{proof}

Suppose $C_i'$ is an arithmetic circuit of size $s_i'$ for $f_i'$ for
$i \in [r]$. We will construct a circuit of size $poly(s_i', n)$ for
$f_i$ efficiently for each $i\in [r]$, which is the crucial part of
our multivariate to bivariate reduction.

The next lemma describes the algorithm crucial to the white-box
reduction.

\begin{lemma}\label{fact-recover}
Given as input a noncommutative arithmetic circuit $C$ for the
polynomial $\varphi(g)\in\F\angle{x,y}$, where $g\in\FX$ is a degree $d$
polynomial, $X=\{x_1,x_2,\ldots,x_n\}$, there is a deterministic
polynomial-time algorithm, running in time $\poly(d,\size(C),n)$ that
computes a noncommutative arithmetic circuit $C'$ for the polynomial
$g$. Furthermore, if $\varphi(g)$ is given by an algebraic branching
program then the algorithm computes an algebraic branching program for
$g$.
\end{lemma}

\begin{proof}  

The proof is based on the idea of evaluating a noncommutative
arithmetic circuit on an automaton (specifically, a substitution
automaton) described in \cite{ARJ15} (see e.g., for related
applications \cite{AJMR17},\cite{AMS10}).
 
Let $g'= \varphi(g)$. Let $g=\sum_m \alpha_m m$ where $m\in X^*$ and
$\alpha_m$ is the coefficient of $m$ in $g$. As noted before, the map
$\varphi$ has the property that $\mon(\varphi(m))\cap
\mon(\varphi(m')= \empty$ for monomials $m \neq m'$ in $X^*$. Moreover
if $m=x_{i_1}x_{i_2}\ldots x_{i_\ell}$ has nonzero coefficient
$\alpha_m$ in $g$ then $g'$ has a monomial $m'=v_{i_1}v_{i_2}\ldots
v_{i_\ell}$ with coefficient $\alpha_m$. Hence, to retrieve an
arithmetic circuit for $g$ from the given circuit $C'$ for $g'$ our
aim is to carry out the following transformation of the polynomial
$g'$ given by the circuit $C'$:

\begin{itemize}
\item Get rid of the monomials of $g'$ containing of all
  $\overline{v_j}\in\overline{T}$ for $j\in[n]$.
\item For each remaining monomial $m'$ of $g'$ substitute $x_i$
  wherever the monomial $v_i$ occurs as substring in $m'$ for $i\in
  [n]$.
\end{itemize}

We will accomplish this transformation by evaluating the circuit $C'$
at suitably chosen matrix substitutions $x\leftarrow M_x$ and
$y\leftarrow M_y$, where $M_x$ and $M_y$ will be $N\times N$ matrices
for polynomially bounded $N$. The resulting evaluation $C'(M_x,M_y)$
will be be an $N\times N$ matrix. A designated entry of this matrix
will contain the polynomial $g$. Clearly, if we can efficiently
compute the claimed matrices $M_x$ and $M_y$ it will yield an
arithmetic circuit $C$ for the polynomial $g$. These matrices $M_x$
and $M_y$ will be obtained as transition matrices of a substitution
automaton that will carry out the above transformation steps on
the polynomial $g'$.

We recall substitution automata in the current context.  A finite
substitution automaton $\mathcal{A}$ is a deterministic finite
automata $\mathcal{A}$ along with a substitution map
\[
\delta: Q \times \{x,y\} \to Q \times (X \cup \mathbb{F})
\]
where $Q$ is a set of states and $X=\{x_1, x_2, \ldots, x_n\}$ are
noncommuting variables. For $i,j \in Q$, $a\in \{x,y\}$, $u \in X \cup
\mathbb{F}$, if $\delta(i,a)=(j,u)$, it means that when automata
$\mathcal{A}$ in state $i$ reads $a$, it replaces $a$ by $u$ and
transitions to state $j$. For each $a \in \{x,y\}$ we can define $|Q|
\times |Q|$ transition matrix $M_a$ such that $M_a (i,j) = u$ if
$\delta(i,a)=(j,u)$ and $0$ otherwise.

With $\delta$ we associate projections $\delta_1:Q \times \{x,y\} \to
Q$ and $\delta_2:Q \times \{x,y\} \to X\cup \mathbb{F}$ defined as
$\delta_1(i,a)=j$ and $\delta_2(i,a)=u$ if $\delta(i,a)=(j,u)$. The
functions $\delta_1$ and $\delta_2$ extend naturally to monomials: For
$w\in \{x,y\}^*$, $\delta_1(i,w)=j$ means the automaton $\mathcal{A}$
goes from state $i$ to $j$ on reading $w$. Let $\tilde{w}_\ell$
denotes length $\ell$ prefix of $w$ and $w_\ell$ denotes $\ell ^{th}$
symbol of $w$ from left. $\delta_2(i,w)=p$ means
\[
p=\prod_{\ell=0}^{|w|-1} \delta_2(\delta_1(i,\tilde{w}_\ell),w_{\ell+1}).
\]
Note that $\delta_2(i,w)$ has the form $\beta \cdot w'$ where $\beta
\in \mathbb{F}, w' \in X^*$. For $\alpha \in \mathbb{F}$ define
$\delta_2(i,\alpha \cdot w)$ as $\alpha \cdot \delta_2(i,w)$.

Let $g'(x,y) = \sum_m \alpha_m m\in\Fxy$. Then, the $(s,t)^{th}$ entry
of the $|Q| \times|Q|$ matrix $g'(M_x,M_y)$ is a polynomial $g\in \FX$
such that
\[
g = \sum_{m \in W_t} \alpha_m \delta_2(s,m),
\]
where $W_t$ is the set of all monomials that take the automaton
$\mathcal{A}$ from state $s$ to state $t$.


Clearly, if $g'$ has an arithmetic circuit of size $s$ then we can
construct an arithmetic circuit of size poly($s,n,|Q|$) for $g$ in
deterministic time poly($s,n,|Q|$).

Turning back to the reduction, consider the input circuit $C$ for $g'=
\varphi(g) \in \Fxy$. We will construct a substitution automaton
$\mathcal{A}$ such that the polynomial $g$ is the $(s,t)^{th}$ entry
of the matrix $g'(M_x, M_y)$.

\paragraph{Description of the Substitution Automata}

As already observed each $v_i$ is of the form $xxw_iyy$, where $w_i$
is a Dyck monomial.  Let $v_i'=xw_iy$ for $i\in [n]$. We can easily
design a deterministic finite automaton $A'$ with $O(mn)$ states such
that the language accepted by $A'$ is precisely the finite set
$\{v_1', v_2', \ldots, v_n'\}$, where $m$ is the length of $v_i$ for $i\in[n]$. Let $\delta'$ denote the transition
function and $Q'$ be the set of states of $A'$, where $q_1$ is the
initial state and $q_{f_i}$ is the final state associated with
acceptance of string $v_i'$ for $i \in [n]$. $A'$ has a tree structure
with root $q_1$ and leaves $q_{f_i}$ for $i\in [n]$, and any root to
leaf path has length exactly $2\ell-2$.  We now define the
substitution automaton $\mathcal{A}$. Its state set is $Q = Q' \cup
\{q_0, q_f, q_r\}$. The transition function $\delta: Q \times \{x,y\}
\to Q \times (X \cup \mathbb{F})$ is defined as follows:
\begin{enumerate}
\item $\delta(q_0,x)=(q_1,1); \delta(q_0,y)=(q_r,0)$.
\item for $q \in Q'\setminus \{q_{f_i} | 1\leq i \leq n \}$.  and
  $a\in \{x,y\}$, let $\delta(q,a) = (\delta'(q,a),1)$.
\item $\delta(q_{f_i},x)=(q_r,0)$; $\delta(q_{f_i},y)=(q_f,x_i)$ for
  each $i\in [n]$.
\item $\delta(q_f,x)=(q_1,1)$ and $\delta(q_f,y)=(q_r,0)$.
\item $\delta(q_r,a)=(q_r,0)$ for $a \in \{x,y\}$.
\end{enumerate} 

The final state of $\mathcal{A}$ is $q_f$. For a monomial $w\in
\{x,y\}^*$, starting at state $q_0$ the automaton $\mathcal{A}$
substitutes all the variables with $1$ as long as it matches with a
prefix of $v_i$ for $i\in [n]$ (given by transitions in 1,2
above). When the monomial matches with $v_i$ for some $i$ (which will
happen while reading symbol $y$ as each string $v_i$ ends with $y$),
$\mathcal{A}$ substitutes $y$ by $x_i$ and moves to state $q_f$. If it
reads $x$ instead of $y$ then $\mathcal{A}$ enters a rejecting state
$q_r$ (given by transition in 3 above). Hence, if $\mathcal{A}$ finds
substring $v_i$ in $w$ it replaces it with $x_i$.  Whenever
$\mathcal{A}$ is in state $q_f$, it means the monomial read so far is
of the form $v_{i_1} v_{i_2} \ldots v_{i_t}$, and it has replaced it
with $x_{i_1}x_{i_2}\ldots x_{i_t}$. If in the state $q_f$ symbol $y$
is encountered, it means the next substring cannot match with a
minimally balanced monomial (as these start with $x$) and the
automaton goes to the rejecting state $q_r$.  If in state $q_f$
variable $x$ is read the automaton goes to state $q_1$ and restarts
the search for a new substring that matches with some $v_i$
(transition in 4 above).

In conclusion $\mathcal{A}$ replaces all the monomials of the form
$v_{i_1} v_{i_2} \ldots v_{i_t}$ by $x_{i_1}x_{i_2}\ldots x_{i_t}$. If
the monomial contains an occurrence of $\overline{v_i}$, or it is not
of the form $v_{i_1} v_{i_2} \ldots v_{i_t}$, then $\mathcal{A}$ zeros
out that monomial by suitably setting an occurrence of $y$ to zero or
enters the reject state $q_r$.\footnote{We can dispense with the
  reject state $q_r$, as suitably setting an occurrence of $y$ to $0$
  would also suffice. We have transitions to the reject state $q_r$
  for exposition.}
 
It follows that the $(q_0,q_f)^{th}$ entry of the $|Q|\times |Q|$
matrix $g'(M_x,M_y)$ is the polynomial $g$, where $g'=\varphi(g)$, and
$M_x,M_y$ are the transition matrices for the substitution automaton
$\mathcal{A}$. This completes the proof.

Finally, if $\varphi(g)$ is given by an algebraic branching program
$P$ then it is easy to see that the above construction with the
substitution automaton $\mathcal{A}$ yields $P(M_x,M_y)$ which is an
algebraic branching program.
\end{proof}




The main theorem of this section, stated below, summarizes the
discussion in this section.
 
\begin{theorem}\label{thm-wb-red}
 In the white-box setting, factorization of multivariate
 noncommutative polynomials into irreducible factors is deterministic
 polynomial-time reducible to factorization of bivariate
 noncommutative polynomials into irreducible factors. More precisely,
 given as input $f\in\FX$ by an arithmetic circuit (respectively,
 algebraic branching program), the problem of computing a complete
 factorization $f=f_1\cdot f_2\cdots f_r$ where each $f_i$ is output
 as an arithmetic circuit (resp.\ algebraic branching program) is
 deterministic polynomial-time reducible to the same problem for
 bivariate polynomials in $\Fxy$.
\end{theorem}

\begin{proof}
  We describe the reduction:
  \begin{enumerate}
  \item Input $f\in\FX$ (as a circuit or ABP).
  \item Transform $f$ to $f'=\varphi(f)\in\Fxy$ as a circuit
    (resp.\ ABP) by the algorithm of Lemma~\ref{lemma-catlan}.
  \item Compute a complete factorization of $f'=f'_1\cdot f'_2\cdots
    f'_r$, where each $f'_i\in\Fxy$ is irreducible and is computed as
    a circuit (resp.\ ABP).
  \item Apply the algorithm of Lemma~\ref{fact-recover} to obtain a
    complete factorization of $f=f_1\cdot f_2\cdots f_r$, where each
    $f_i$ is irreducible and is output as circuit (resp.\ ABP).
\end{enumerate}

The correctness of the reduction and its polynomial time bound follow
from Lemmas \ref{lemma-inert}, \ref{lemma-catlan} and
\ref{fact-recover}.
\end{proof}

\begin{remark}
  We note that in the case $\F$ is the field $\Q$ (of rationals), we
  need to take into account the bit complexity of the rational numbers
  involved and argue that the reduction is still polynomial time
  computable.  The main point to note here is that the reduction
  guarantees the size of the factor $f_i$ is polynomially bounded in
  the size of $g_i, 1\le i\le r$, where the size of $g_i$ includes the
  sizes of any rational numbers that might be involved in the
  description of the arithmetic circuit (or ABP) for $g_i$.
\end{remark}

\begin{remark}
We note here that the ring $\FX$ is not a unique factorization domain.
That is, a polynomial $f \in \FX$ may have, in general, multiple
factorizations into irreducibles \cite{Cohnfir}. A standard example is
the polynomial $x+xyx$ which factorizes as $x(1+yx)$ as well as
$(1+xy)x$, where $x, y, 1+yx, 1+xy$ are irreducible. As the map
$\varphi$ is an injective homomorphism, there is a 1-1 correspondence
between factorizations of $\varphi(f)$ and factorizations of $f$. More
specifically, our reduction takes as input any complete factorization
$\varphi(f)= f_1' f_2' \ldots f_r'$ and computes the corresponding
complete factorization $f=f_1 f_2 \ldots f_r$ of $f$.
\end{remark}

\begin{remark}  
We note that the embedding $\varphi$ does not preserve
sparsity\footnote{The sparsity of a polynomial $f$ is the number of
  monomials in $\mon(f)$.} of the polynomial $f$. More precisely, if
the sparsity of the $n$-variate degree $d$ polynomial $f$ is $s$ then the
sparsity of the bivariate polynomial $\varphi(f)$ is $O(2^d s)$. Thus,
using this embedding map we do not get a reduction from sparse
$n$-variate degree $d$ polynomial factorization to sparse bivariate polynomial
factorization, where $s,d$ are allowed to be part of the running
time. This problem remains unanswered.
\end{remark}

\subsection{Black-box reduction}

The reduction in the black-box case is essentially identical. The only
point to note, which is easy to see, is the that analogue of
Lemma~\ref{fact-recover} holds in the black-box setting. We state that
below. We recall what a black-box means in the noncommutative setting.

\begin{definition}\label{nc-bb-def}  
  A noncommutative polynomial $f\in\FX$ given by \emph{black-box}
  essentially means we can evaluate $f$ at any matrix substitution
  $x_i\leftarrow M_i$, $M_i\in\F^{N\times N}$, where the cost of each
  evaluation is the matrix dimension $N$.
\end{definition}  

In the black-box setting, suppose we have an efficient algorithm for
bivariate noncommutative polynomial factorization of degree $D$
polynomials $g\in\F\angle{x,y}$, where the algorithm takes a black-box
for $g$ and outputs black-boxes for the irreducible factors of some
factorization of $g$ in time $\poly(D)$. Then, given a black-box for a
degree $D$ $n$-variate polynomial $f\in\FX$ as input, we require that
the reduction transforms it into a black-box of a bivariate polynomial
$g\in\F\angle{x,y}$, and from the output black-boxes of $g$'s
irreducible factors, the reduction has to efficiently recover
black-boxes for the corresponding irreducible factors of $f$.

\begin{lemma}\label{fact-recover-bb}
Given as input a black-box for the polynomial
$\varphi(g)\in\F\{x,y\}$, where $g\in\FX$ is a degree $d$ polynomial,
$X=\{x_1,x_2,\ldots,x_n\}$, with matrix substitutions for $x$ and $y$
computed in deterministic polynomial-time time we can obtain a
black-box for the polynomial $g\in\FX$.
\end{lemma}

\begin{proof}
  The proof of Lemma~\ref{fact-recover} already implies this because
  the matrices $M_x$ and $M_y$ described there do not require
  $\varphi(g)$ to be given in white-box as circuit or ABP. Thus, the
  black-box for $\varphi(g)$ yields a black-box for $g$ by accessing
  the $(q_0,q_f)^{th}$ entry of the matrix output
  $\varphi(g)(M_x,M_y)$. 
\end{proof}

As a consequence we obtain the claimed reduction from multivariate
factorization to bivariate factorization in the black-box setting as
well.

\begin{theorem}\label{thm-bb-red}
  The problem of computing a complete factorization of $f\in\FX$ given
  by black-box is deterministic polynomial-time reducible to the
  problem of black-box computation of a complete factorization of
  polynomials in $\Fxy$.
\end{theorem}  

\begin{proof}
  Given a black-box for $f$ we obtain a black-box for $\varphi(f)$
  applying Lemma~\ref{lemma-catlan}. Then, given a complete
  factorization
\[
\varphi(f) = f'_1\cdot f'_2\cdots f'_r,
\]
where each factor $f'_i$ is output by a black-box for it, by
Lemma~\ref{fact-recover-bb} we can obtain black-boxes for each $f_i$.
This yields a complete factorization $f=f_1\cdot f_2\cdots f_r$ of $f$
where the factors are given by black-box.
\end{proof}


\section{Factorizing $4\times 4$ linear matrices over $\Q$}

We have shown in Section~\ref{sec-multi-bi} that multivariate noncommutative
polynomial factorization is efficiently reducible to the bivariate
case. Suppose $f\in\F\angle{x,y}$ is a bivariate polynomial given by a
formula of size $s$. Applying Higman linearization \cite{Cohnfir}, as
done in \cite{AJ22}, we can transform the problem to factorization of
bivariate linear matrices $A_0+A_1x+A_2y$, where the matrices have
size bounded by $2s$. In \cite{AJ22} the problem of factorizing an
$n$-variate polynomial $f\in\FX$ given by a formula was solved in two
steps when $\F$ is a finite field: (i)~ Transform $f$ to a linear
matrix $L$ and factorize $L$ into irreducible factors by reducing it
to the common invariant subspace problem, and (ii)~ extract the
factors of $f$ from the factors of $L$.  This approach does not work
for $\F=\Q$ because the common invariant subspace problem for matrices
over $\Q$ is shown by Ronyai \cite{Ronyai} to be at least as hard as
factoring square-free integers.

In this section we show that even for $4\times 4$ bivariate linear
matrices factorization remains at least as hard as factoring
square-free integers. Thus, efficient polynomial factorization over
$\Q$ remains elusive even for bivariate polynomials. The proof
is based on Ronyai's aforementioned result.

\begin{definition}[generalized quaternion algebra]
  Let $\alpha,\beta\in Q$ be nonzero rationals. The generalized
  quaternion algebra $H(\alpha,\beta)$ is the 4-dimensional algebra
  over $\Q$ generated by elements $1, u, v, uv$ where the rules for
  multiplication in $H(\alpha,\beta)$ are given by $u^2=\alpha$,
  $v^2=\beta$, and $uv=-vu$.
\end{definition}

A \emph{simple} algebra $\mathcal{A}$ over a field $\F$ is an algebra
that has no nontrivial two-sided ideal. The \emph{center} $C$ of
algebra $\mathcal{A}$ is the subalgebra consisting of all elements of
$\mathcal{A}$ that commute with every element of $\mathcal{A}$.

\begin{fact}
For any nonzero $\alpha,\beta\in\Q$, the algebra $H(\alpha,\beta)$ is
a simple algebra with center $\Q$.
\end{fact}  

Furthermore, it follows from general theory \cite[Chapter 1.6]{Pierce}
that

\begin{fact}
  The algebra $H(\alpha,\beta)$ is either a division algebra (which
  means no zero divisors in it) or is isomorphic to the algebra of
  $2\times 2$ matrices over $\Q$ (which means it has zero divisors).
\end{fact}

The 4-dimensional algebra $H(\alpha,\beta)$ can be represented as an
algebra of $4\times 4$ matrices over $\Q$, which is the \emph{ regular
  representation}. It is easy to see that the matrix corresponding to
$1$ is $I_4$, and the matrices $M_u$ and $M_v$ corresponding to $u$ and
$v$ are
\begin{equation}\label{mu}
M_u  = 
\begin{bmatrix} 
  0 & 1 & 0 & 0 \\
  \alpha & 0 & 0 & 0 \\
   0 & 0 &0 & 1 \\
   0 & 0 & \alpha & 0 
\end{bmatrix} 
\end{equation}

\begin{equation}\label{mv}
M_v  =  \begin{bmatrix} 
  0 & 0 & 1 & 0 \\
  0 & 0 & 0 & -1 \\
   \beta & 0 &0 & 0 \\
   0 & -\beta & 0 & 0 
\end{bmatrix}
\end{equation}

We next observe that factorizing $4\times 4$ bivariate linear matrices
is at least as hard as finding zero divisors in generalized quaternion
algebras.

\begin{theorem}
  Finding zero divisors in an input quaternion algebra
  $H(\alpha,\beta)$ is polynomial-time reducible to factorizing
  $4\times 4$ bivariate linear matrices $A_0+A_1x + A_2y$, where each
  scalar matrix $A_i$ is in $\M_4(\Q)$.
\end{theorem}

\begin{proof}
  Let $H(\alpha,\beta)$ be the given generalized quaternion algebra. Then
  \[
  H(\alpha,\beta) = \{a_o + a_1 u + a_2 v + a_3uv\mid a_i\in\Q\},
  \]
  where $u^2=\alpha$, $v^2=\beta$, and $uv=-vu$ defines the algebra
  multiplication.

  It is well-known (see e.g.\ Pierce's book \cite[Chapter
    1.6]{Pierce}) that the algebra $H(\alpha,\beta)$ is
    \emph{simple} (that is, it has no nontrivial 2-sided ideals) with
    center $\Q$. Furthermore, it is either a \emph{division algebra}
    (which means there are no zero divisors in it) or it is isomorphic
    to the algebra $\M_2(\Q)$ of $2\times 2$ matrices over $\Q$ (which
    has zero divisors).
  
We now consider factorizations of the $4\times 4$ linear matrix $I_4
+ M_u x + M_v y$, where matrices $M_u$ and $M_v$ are defined in
Equations \ref{mu} and \ref{mv}.  

  \begin{claim}
    The linear matrix $I_4+M_u x + M_v y$ is irreducible if and only if
    the quaternion algebra is a division algebra.
   \end{claim} 

  \begin{proofof}
    
Suppose the linear matrix $L=I_4 + M_u x + M_v y$ has a nontrivial factorization
\[
L=I_4 + M_ux+M_vy = FG.
\]
That means neither $F$ nor $G$ is a scalar matrix. By a theorem of
Cohn \cite[Theorem 5.8.8]{Cohnfir}, there are invertible scalar matrices
$P$ and $Q$ in $\M_4(\Q)$ such that
\begin{equation}\label{blockfact}
PLQ =
\begin{bmatrix}
  A & 0\\
  D & B
\end{bmatrix}.
\end{equation}

\begin{remark}
To apply Cohn's theorem we need to have matrix $L$ to be monic (that is the matrix $[M_u ~|~ M_v]$ has full row rank and the matrix $[M_u ^T~ | ~M_v ^T ]^T$ has full column rank). The monicity is ensured for $L$ as matrices $M_u$ and $M_v$ are full rank matrices.
\end{remark}

Putting $x=y=0$ we observe that
\[
PQ =\begin{bmatrix}
  A_0 & 0\\
  D_0 & B_0
\end{bmatrix},
\]
where $A_0, B_0$ and $D_0$ are scalar matrices. As $P$ and $Q$ are
invertible, it following that both $A_0$ and $B_0$ are invertible
matrices. Hence we have
\[
PLP^{-1}=\begin{bmatrix}
  A & 0\\
  D & B
\end{bmatrix}\cdot
\begin{bmatrix}
  A_0 & 0\\
  D_0 & B_0
\end{bmatrix}^{-1}
=\begin{bmatrix}
  A' & 0\\
  D' & B'
\end{bmatrix},
\]
where $A', B'$ and $D'$ are also linear matrices. We now recall that
the matrices $I_4, M_u$ and $M_v$ are the matrix representation of the
elements $1, u, v\in H(\alpha,\beta)$ w.r.t.\ the basis $\{1,u,v,uv\}$
is the basis of $H(\alpha,\beta)$. Treating $P$ as a basis change
matrix, the above equation yields a new basis $\{w_1,w_2,w_3,w_4\}$ of
$H(\alpha,\beta)$. Let $\dim(A')=k$. Then $1\le \dim(A')\le 3$ and the
vectors $w_1,\ldots,w_k$ spans a $k$-dimensional subspace $W\subset
H(\alpha,\beta)$ that is a common invariant subspace for the matrices
$I_4, M_u, M_v$ and $M_{uv}$. In other words, the subspace $W$ is
preserved under left multiplication by $u$ and $v$.  We can assume,
without loss of generality, that $w_1\ne 1$: if $k>1$ then clearly we
can assume this. If $k=1$ notice that $w_1=1$ is impossible because
the subspace $W$ is not preserved under left multiplication by $u$ or
$v$. Then the four elements $w_1, uw_1, vw_1, uvw_1$ are all in $W$
and hence linearly dependent. Thus for some nontrivial linear
combination
\[
\gamma_0w_1+\gamma_1 uw_1 +\gamma_2 vw_1 +\gamma_3 uvw_1 = 0.
\]
which means $(\gamma_0+\gamma_1 u +\gamma_2 v +\gamma_3 uv)\times w_1
= 0$. Hence $w_1$ is a zero divisor in $H(\alpha,\beta)$.

Conversely, if $z\in H(\alpha,\beta)$ is a zero divisor then the we can see that
the left ideal
\[
J = \{xz\mid x\in H(\alpha,\beta)\}
\]
is a proper subspace of $H(\alpha,\beta)$ that is invariant under
$M_u$ and $M_v$. Then, applying Cohn's theorem \cite[Theorem
  5.8.8]{Cohnfir}, we can obtain invertible scalar matrices $P$ and $Q$
such that Equation~\ref{blockfact} holds which yields the factorization
\[
PLQ =
\begin{bmatrix}
  A & 0\\
  D & B
\end{bmatrix} =\begin{bmatrix}
  A & 0\\
  0 & I
\end{bmatrix} \cdot
\begin{bmatrix}
  I & 0\\
  D & I
\end{bmatrix} \cdot
\begin{bmatrix}
  I & 0\\
  0 & B
\end{bmatrix}.
\]
\end{proofof}  

To complete the reduction, notice that if $I_4+M_ux+M_vy$ is
irreducible then $H(\alpha,\beta)$ is a division algebra. On the other
hand, if we are given a nontrivial factorization $I_4+M_ux+M_vy = FG$
then, analyzing the proof of Cohn's theorem \cite[Theorem 5.8.8]{Cohnfir}
(also see \cite{AJ22} for details), by suitable row and column
operations we can compute in polynomial time the invertible scalar
matrices $P$ and $Q$ from the factors $F$ and $G$. Hence, by the proof
of the above claim, we can efficiently compute a zero divisor $w_1$ in
$H(\alpha,\beta)$. 
\end{proof}

As finding zero-divisors in the quaternion algebra $H(\alpha,\beta)$
is known to be at least as hard as square-free integer factorization
\cite{Ronyai} we have the following.

\begin{corollary}
  Factorizing $4\times 4$ bivariate linear matrices over $\Q$ is at
  least as hard as factorizing square-free integers.
\end{corollary}

\section{Factorizing $3\times 3$ linear matrices over $\Q$}

In this section we present a deterministic polynomial-time algorithm
for factorization of $3\times 3$ multivariate linear matrices over
$\Q$. We start with a simple observation about linear matrix
factorization in general.

\begin{lemma}
Suppose $L=I_d+\sum_{i=1}^nA_ix_i$ is a linear matrix where each $A_i,
0\le i\le d$ is a $d\times d$ matrix over $\Q$. Then $L$ is
irreducible if the characteristic polynomial of $A_i$ is irreducible
over $\Q$ for any $i$.
\end{lemma}

\begin{proof}
  For if $L$ is reducible then there is an invertible scalar matrix $P$ such
  that
  \[
  PLP^{-1}=
  \begin{bmatrix}
    A & 0\\
    D & B
  \end{bmatrix},
  \]
  which implies that
  \[
  PA_iP^{-1}=
  \begin{bmatrix}
    A'_i & 0\\
    D'_i & B'_i
  \end{bmatrix},
  \]
  for scalar matrices $A'_i, B'_i$, and $D'_i$.  Thus, the
  characteristic polynomial of $A_i$ is the product of the
  characteristic polynomials of $A'_i$ and $B'_i$ which is a
  nontrivial factorization.
\end{proof}

\begin{theorem}
There is a deterministic polynomial-time algorithm for factorization
of $3\times 3$ multivariate linear matrices over $\Q$.
\end{theorem}

\begin{proof}
We will first consider linear matrices of the form $L=I_3 +
\sum_{i=1}^n A_ix_i$, where each $A_i\in \M_3(\Q)$ and the $x_i$ are
noncommuting variables. The algorithm computes a complete
factorization of $L$ into (at most three) irreducible linear matrix
factors. By Cohn's theorem \cite[Theorem 5.8.8]{Cohnfir}, either $L$ is
irreducible or there is an invertible scalar matrix $P$ such that
\[
PLP^{-1} =
\begin{bmatrix}
  A & 0\\
  D & B
\end{bmatrix}.
\]

Either $A$ or $B$ is a $1\times 1$ matrix. If $A$ is a $1\times 1$
matrix then corresponding to it there is a 1-dimensional common
invariant subspace spanned by a vector, say $v$, for the matrices
$A_i, 1\le i\le n$. More precisely, the row vector $v^T$ is an
eigenvector for each matrix $A_i$, and $v^T A_i = \lambda_i v^T$ where
$\lambda_i\in\Q$ is the corresponding eigenvalue of matrix $A_i$ for
each $i$. Likewise, if $B$ is a $1\times 1$ matrix then there is a
corresponding 1-dimensional common invariant subspace spanned by a
(column) vector $u$ such that $A_i u =\mu_i u$ for eigenvalues $\mu_i$
of $A_i$. In either case, the common eigenspace is easy to compute
from the characteristic polynomial of say $A_1$ and then verifying
that it is an eigenspace for the remaining $A_i$ as well. This will
yield the factorization
\[
PLP^{-1}=\begin{bmatrix}
  A & 0\\
  0 & I
\end{bmatrix} \cdot
\begin{bmatrix}
  I & 0\\
  D & I
\end{bmatrix} \cdot
\begin{bmatrix}
  I & 0\\
  0 & B
\end{bmatrix},
\]
where $B$ is a $2\times 2$ linear matrix. The problem now reduces to
factorizing the linear matrix $B= I_2 + \sum_{i=1}^n B_ix_i$, where
$B_i\in\M_2(\Q)$. A simple case analysis discussed below yields a
polynomial-time algorithm for factorization of $B$.

\begin{enumerate}
\item If the characteristic polynomial of any $B_i$ is irreducible over
  $\Q$ then the linear matrix $B$ is clearly irreducible.
\item Some $B_i$ has two distinct eigenvalues $\lambda\ne \lambda'\in
  \Q$ then the corresponding eigenspaces are 1-dimensional, spanned by
  their eigenvectors $u\ne u'$. Then either $u$ or $u'$ has to be an
  eigenvector for every $B_j$ (otherwise $B$ is irreducible), in which
  case we have a factorization of $B$.
\item Suppose each $B_i$ has only one eigenvalue $\lambda_i$. Then,
  by linear algebra, after a basis change $B_i$ is either of the form
  \[
  \begin{bmatrix}
    \lambda_i & 1\\
    0  & \lambda_i
   \end{bmatrix} 
  \]
  in which case the eigenspace is 1-dimensional with eigenvector $(1 0)^T$.
  We can check if this eigenspace is invariant for each $B_j$ or not as
  before. Otherwise, after basis change each
\[
B_i=
\begin{bmatrix}
    \lambda_i & \\
    0  & \lambda_i
   \end{bmatrix} 
\]
which means $B_i=\lambda_i I_2$ for each $i$ and the factorization of
$B$ is given by
\[
B=\begin{bmatrix}
  1+\sum_{i=1}^n\lambda_ix_i & 0\\
  0 & 1
\end{bmatrix} \cdot
\begin{bmatrix}
  1 & 0\\
  0 & 1+\sum_{i=1}^n\lambda_ix_i
\end{bmatrix}
\]
\end{enumerate}
\end{proof}

\bibliographystyle{plain}

\bibliography{references}

\end{document}